\documentclass[fleqn,10pt]{wlscirep}
\usepackage[utf8]{inputenc}
\usepackage[T1]{fontenc}
\title{Virtual Frame Technique: \\ Ultrafast Imaging with Any Camera}

\author[1]{Sam Dillavou}
\author[2]{Shmuel M Rubinstein}
\author[3,*]{John M Kolinski}
\affil[1]{Harvard University FAS, Cambridge, MA 02138, USA, dillavou@g.harvard.edu}
\affil[2]{Harvard University SEAS, Cambridge, MA 02138, USA, shmuel@seas.harvard.edu}
\affil[3]{EPFL IGM, Lausanne, VD 1015, Switzerland, john.kolinski@epfl.ch}
\affil[*]{Corresponding Author}

\begin{abstract}
Many phenomena of interest in nature and industry occur rapidly and are difficult and cost-prohibitive to visualize properly without specialized cameras. Here we describe in detail the Virtual Frame Technique (VFT), a simple, useful, and accessible form of compressed sensing that increases the frame acquisition rate of any camera by several orders of magnitude by leveraging its dynamic range. VFT is a powerful tool for capturing rapid phenomenon where the dynamics facilitate a transition between two states, and are thus binary. The advantages of VFT are demonstrated by examining such dynamics in five physical processes at unprecedented rates and spatial resolution: fracture of an elastic solid, wetting of a solid surface, rapid fingerprint reading, peeling of adhesive tape, and impact of an elastic hemisphere on a hard surface. We show that the performance of the VFT exceeds that of any commercial high speed camera not only in rate of imaging but also in field of view, achieving a 65MHz frame rate at 4MPx resolution. Finally, we discuss the performance of the VFT with several commercially available conventional and high-speed cameras. In principle, modern cell phones can achieve imaging rates of over a million frames per second using the VFT.
\end{abstract}

\begin{document}

\flushbottom
\maketitle
\thispagestyle{empty}

\section*{Introduction}
Nature is rife with phenomena that occur faster than our senses can capture. As a result, progress in imaging technology often yields new physical insight \cite{Edgerton:1937bl}. Because the rate of data transfer is capped by hardware technology, increasing frame acquisition rate is conventionally achieved by reducing image resolution, sometimes to the extreme of recording only a few lines of pixels\cite{Rubinstein:2004ek}, or even just a few single pixels\cite{Schleip:1988ut}. However, for many rapid phenomena, resolving the 2D dynamics is indispensible. For example, the shape of a crack front reveals the type of fracture underway \cite{Long:2011ey}. Often, processes with rapid, fully two-dimensional dynamics consist of a transition between two states, such as fractured and un-fractured, or wet and non-wet \cite{Kolinski:2012fx}. For these dynamics, the standard use of a camera is extremely inefficient; a 16-bit camera has over 65,000 grayscale values, while only two are necessary. In principle, by sacrificing this (unnecessary) bit-depth, one may may boost the frame rate of any camera by several orders of magnitude while retaining full spatial resolution.

Here we describe a method of imaging dynamic processes in full spatial and enhanced temporal resolution that exploits the dynamic range of the imaging sensor. Binary dynamics are integrated over each exposure time: a time-lapse at ultra-high speeds. A single image, referred to here as a compressed frame stack (CFS), can then be deconvolved into several thousands of `virtual frames' using thresholding. In this way, ultra-fast frame rates are attained using the Virtual Frame Technique (VFT), while the avoiding trade-off between spatial and temporal resolution typical of conventional high-speed imaging. We directly compare the VFT with a standard fast camera by recording a rapid fracture simultaneously using both methods. The VFT results in a higher resolution, larger field of view, and faster frame rate, while faithfully reproducing the images recorded with the fast camera. The VFT allows the imaging of a wetting front during rapid droplet impact in 2D, achieving frame rates of nearly 8MHz. Using a pulsed light source to shorten exposure time, three additional examples of 2D dynamics are recorded at unprecendented speed and resolution, achieving a peak frame rate above 65MHz at 4MPx. Finally, the frame rates and resolutions attainable using standard operation vs the VFT are discussed for several commercially available cameras.

\begin{figure}[ht] %%%%%%%%%%%%%%%%%%%%%%%%%%%%%%%%%%%%%%
\centering 
\includegraphics[width = 1\textwidth]{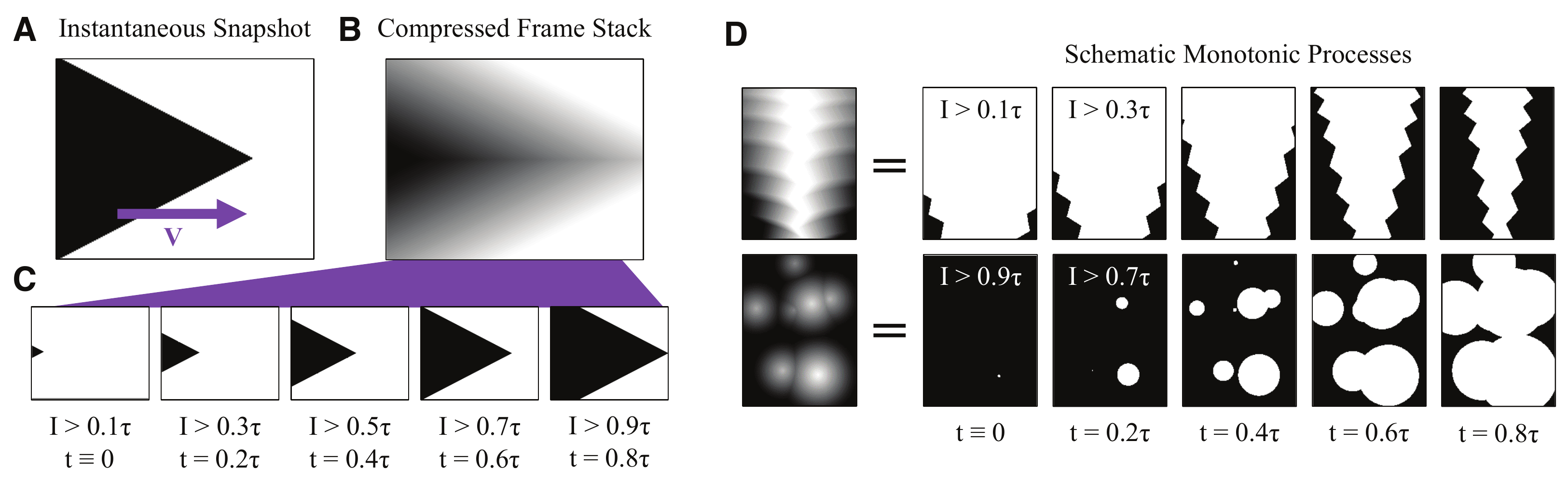}
\caption{\textbf{Virtual Frame Technique: Schematic Demonstration of Working Principle} The dynamics of a v-shaped front propagating at velocity $V$ and traversing the field of view of a camera is shown schematically in both an instantaneous snapshot (a) and a resulting camera image (b), referred to as a Compressed Frame Stack (CFS), recorded with exposure time $\tau$. The grayscale intensity $I(x)$ recorded in the CFS is a convolution of the spatio-temporal dynamics of the propagating front, and thus appears blurred. (c) The dynamics are both binary and monotonic, therefore the instantaneous front position can be obtained by deconvolving the compressed frame stack, $I(x)$. By way of example, five virtual frames are reconstructed by thresholding $I(x)$; these virtual frames correspond to the indicated fractional exposure times $t$. (d) Two additional hypothetical processes that satisfy the binary and monotonic requirements are shown. The relevant CFS (left) is deconstructed into its constituent virtual frames (right) through thresholding. Note that the dynamics may be evolving in all directions, and from light to dark or vice versa.
\label{1}}
\end{figure} %%%%%%%%%%%%%%%%%%%%%%%%%%%%%%%%%%%%%%%

The Virtual Frame Technique boosts frame rate with no loss in spatial resolution by convolving many `virtual frames' into a single image. Consider a hypothetical monotonic process, like the movement of a V-shaped front shown in Fig \ref{1}. The region of interest is illuminated uniformly such that the instantaneous light intensity reaching each pixel is either value $a$ or value $b$ at all times, $i(x,y,t) = {a,b}$ for any $x,y,t$, as shown for $a=0$, $b=1$ in Fig \ref{1}a. Under these conditions, a dynamic process recorded with a finite exposure time $\tau$ will appear blurred, as shown in Fig \ref{1}b. The gray-scale value of a pixel, $I(x,y)$, is a measure of integrated light intensity 
\begin{equation}
I(x,y) = \int_{0}^{\tau} i(x,y,t) dt
\label{eq1}
\end{equation}
for an image taken at time $t=0$. For a monotonic process, each pixel will transition from light to dark at a single time, $t_t(x,y)$\footnote{$t_t = 0$ corresponds to a pixel dark when the exposure begins, and $t_t = \tau$ corresponds to a pixel that remains illuminated when the exposure is complete}. As a result, intensity uniquely maps to transition time, and we can re-write equation \ref{eq1} as:
\begin{equation}
I(x,y) = a(t_t(x,y)) + b(\tau-t_t(x,y)) \quad \rightarrow \quad t_t(I) = (b\tau-I)/(b-a)
\label{transTime}
\end{equation}

Gray scale of the image encodes temporal information; all pixels with a gray value at the threshold $I_0$ represent the location of the front at the corresponding time $t_t(I_0)$. As a result, one can create `virtual frames' by thresholding the blurred image, as shown in Fig \ref{1}c. These virtual frames are instantaneous snapshots of the process at a given time within the exposure, and thousands or more may be stored in a single image (CFS).

The VFT offers significant enhancement of temporal resolution without compromising spatial resolution of dynamic processes. The principle advantage arises from the way the VFT exploits the large bit-depth of typical digital camera sensors: for 16-bit sensors, over 65,000 discrete positions can be recorded in a single exposure, leading to an enhancement of the frame rate by a factor in the tens of thousands. Ideally, virtual frames are limited in number only by the bit depth of the camera, which represents the number of discrete grayscale values, and thus time steps that a single image can contain. The maximum frame rate achievable through this method is
\begin{equation}
fps_{max} = \|b-a\|*2^{\#bits}/\tau \equiv \beta/\tau
\label{fpsMAX}
\end{equation}
where $\beta$ is the `boost' in frame rate, the number of virtual frames per frame, and $\tau$ once again as the exposure time. In this formula $a=1$ or $b=1$ corresponds to exact saturation of a pixel. The frame rate is reduced, however, by the noise inherent to the camera's sensor: whereas for sCMOS sensors the readout noise approaches 1 electron rms, a typical EMCCD will exceed this performance at the lowest light levels, and record nearly perfect dark signals. By contrast, a typical high speed camera sensor has several electrons of readout noise and a lower effective bit-depth, ultimately reducing the attainable $\beta$. In this way the sensor type and its noise characteristics do enter into the performance capabilities of the VFT. Furthermore, noise in the lighting, resulting in variations in $a$ and $b$ across the image, may increase the minimally resolved timestep. However, smooth spatial variation in the illumination can be accounted for by obtaining images of $a(x,y)$ and $b(x,y)$ and simply evaluating equation \ref{transTime} for each pixel.

\begin{figure}[ht] %%%%%%%%%%%%%%%%%%%%%%%%%%%%%%%%%%%%%%
\includegraphics[width = 1\textwidth]{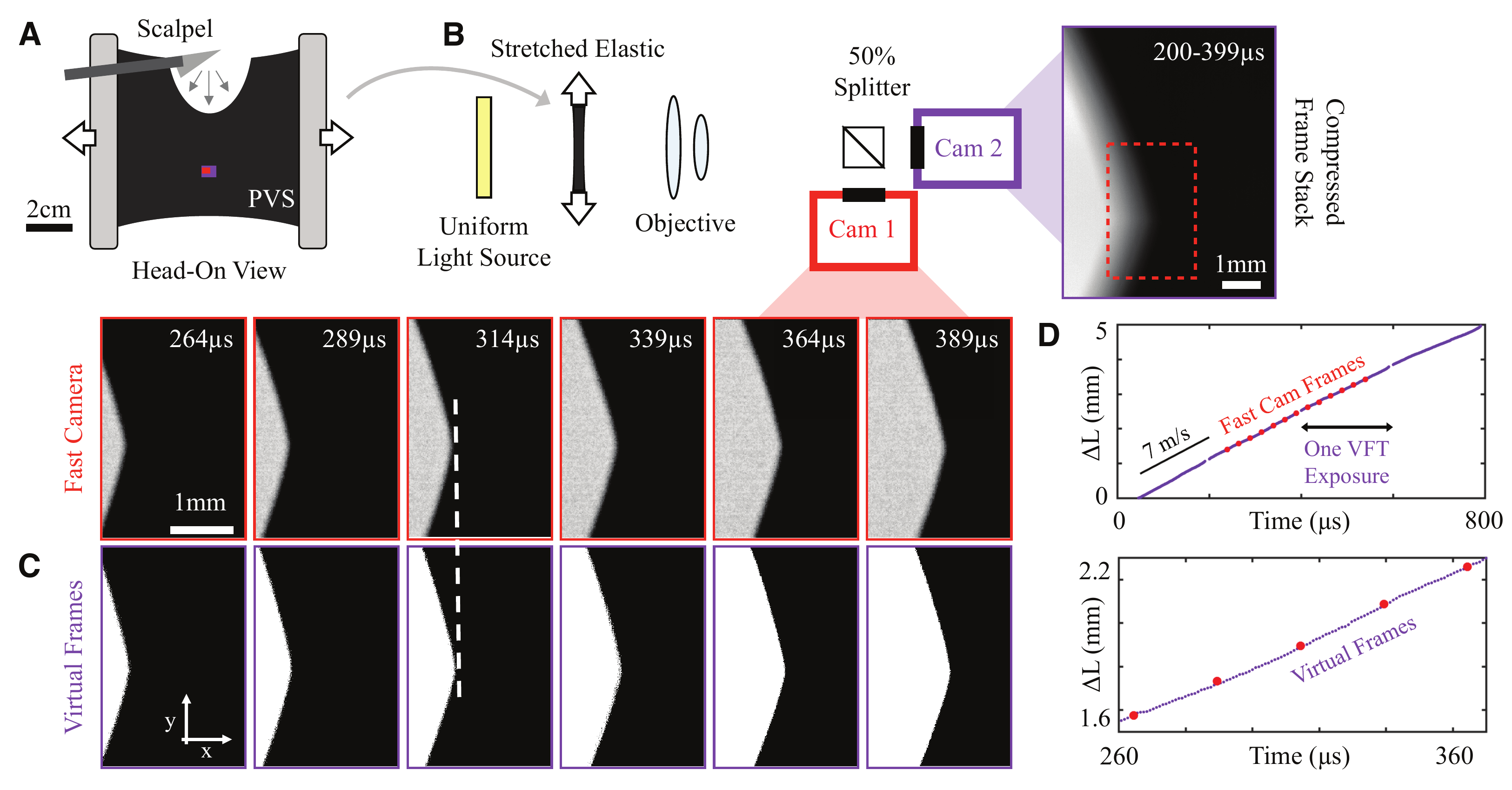}
\caption{\textbf{VFT and Fast Camera Comparison Using Fracture Dynamics}
(a) Experimental setup for measurement of a propagating crack tip. A dynamic fracture is initiated in a strained elastomer made of PVS (polyvinyl siloxane). (b) Background lighting and sample opacity were tuned such that the process appears binary at any instant. Two cameras simultaneously record the dynamics such that the crack moves from left to right across both fields of view. Camera one (red) films at 40KHz with a resolution of 320x208 pixels with a total of 60 kilopixels. Six contrast enhanced images from camera one are shown outlined in red. Camera two (purple) films at 5KHz with a resolution of 1280x1000 pixels with a total of nearly 1.3 megapixels; a compressed frame stack is shown outlined in purple at right. The field of view of camera one is superimposed upon the raw image from camera two with a red dashed box. (c) Virtual frames reconstructed from the raw image in (b) are cropped to match the field of view of camera one. The fractional exposure times are chosen such that they correspond to the images recorded by the fast camera in (b). (d) Top: The crack tip location ($\Delta L$) is measured using both the images of camera one and the virtual frames of camera two. The effective frame rate achieved using VFT is 1MHz, corresponding to $\beta = 200$. Because the field of view of the virtual frames is larger than the fast camera frames, the VFT camera (cam two) tracks the crack tip for nearly three times as long as the camera simply filming (cam one). Bottom: A sub-set of the data plotted on a smaller scale highlights the enhanced temporal resolution of the VFT.
\label{2}}
\end{figure} %%%%%%%%%%%%%%%%%%%%%%%%%%%%%%%%%%%%%%%

\section*{Results}

\subsection*{Proof of Principle: Virtual and Actual Frames of Fracture}

Material failure is a highly dynamic process that often occurs at rates that approach the material sound speeds\cite{freund}. Imaging data have been used in a similar optical configuration to measure the crack tip opening displacement (CTOD), whose curvature\cite{Livne:2008bn, Boue:2015kpa, Goldman:2010kj} and scaling\cite{Kolvin:2018wb} carry essential information about the state of stress at the crack tip, and can be used to measure the stress intensity factor. The VFT enables us to use the entire imaging sensor at high speed, capturing the CTOD in higher resolution in both time and space. For dynamic fracture processes, the VFT can record the crack tip trajectory at a high rate over a large area of interest, enabling detailed studies of the crack's equation of motion\cite{Goldman:2010kj} and the instabilities of a propagating crack\cite{Fineberg:1991jj, Sharon:1996dn}. The binary and monotonic criteria required of the VFT can also be realized using other fracture visualization techniques than the one used here, such as viewing the advancing dynamic fracture front through the sample\cite{Kolvin:2015eb, Kolvin:2017bf}. While imaging clearly provides tremendous insight into the dynamics of rapid fracture, these dynamics are extremely difficult to image even with state-of-the-art high speed cameras, leaving room for significant advancement using the VFT. 

Using VFT simultaneously with conventional high-speed imaging, we record the dynamics of a crack propagating in a soft elastomer. A 1.5 mm thick sheet of polyvinylsiloxane (PVS) is loaded in tension, and a crack is introduced on the sample's edge which spontaneously accelerates through the sample, as shown schematically in Fig. \ref{2}a. An identical projection of the test section is directed onto the two imaging sensors simultaneously using a beam splitter, as shown in Fig. \ref{2}b. Six consecutive images from the high-speed camera are shown atop six virtual frames reconstructed from the compressed frame stack (CFS); the virtual frames reproduce exactly the same front geometry as the instantaneous high-speed images, as can be readily seen in Fig. \ref{2}b and \ref{2}c. The virtual frames were significantly cropped to match the reduced field of view of the high speed camera, and faithfully reproduce the dynamics recorded with the conventional fast camera even on the smallest scales. Using these virtual and fast camera images, the location of the crack tip ($\Delta L$) is calculated as a function of time, as shown in Fig \ref{2}d. Here, the virtual frame rate used is approximately 1 MHz ($\beta = 200$). There is excellent agreement between the two methods, confirming the merit of the VFT. Furthermore, the VFT creates frames 25 times faster\footnote{Here we have limited the virtual frame rate such that each frame corresponds to approximately one pixel of movement of the front. A higher frame rate is achievable with this data, but not especially useful for the purposes of measuring the front position.}, nearly triples the tracking time, and simultaneously increases the field of view, resulting in significant enhancement of data quality.

\begin{figure}[ht] %%%%%%%%%%%%%%%%%%%%%%%%%%%%%%%%%%%%%%
\includegraphics[width = 1\textwidth]{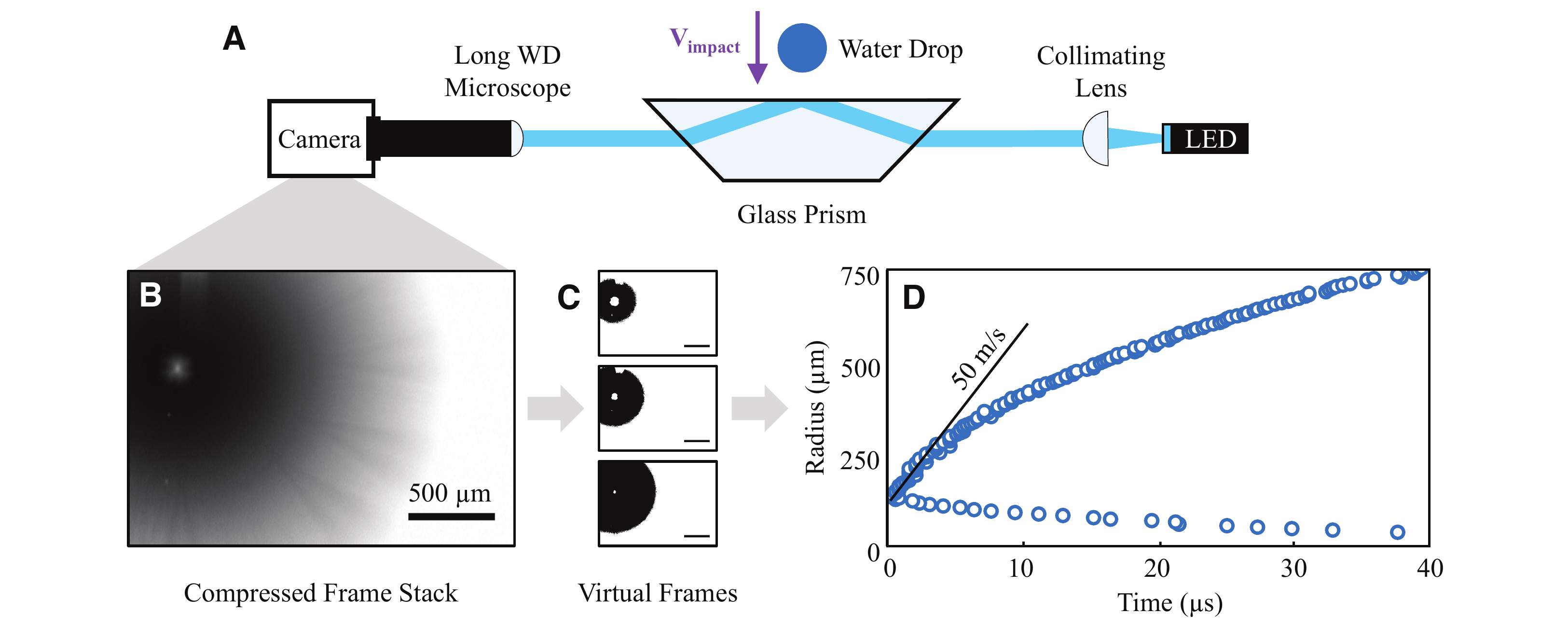}
\caption{\textbf{Wetting Front Propagation Recorded with the VFT}
a) Experimental and imaging setup. A water droplet impacts a glass prism at velocity $V_{impact}$. The surface is illuminated in total internal reflection using a collimated LED. The reflected light is imaged upon the camera’s sensor using a long-working distance microscope objective. Because a wetted surface is no longer totally internally reflecting, pixels sampling the wetted area appear black. b) A typical compressed frame stack of the TIR signal with $V_{impact}$ = 3.5m/s and exposure time $\tau$ = 100$\mu$s. c) Virtual frames are created by thresholding the compressed frame stack. The small divot at the top of the circular front is an optical abberation. d) Contact radius of many virtual frames vs time. Note that the contact initially spreads at a rate exceeding 50 m/s. The inward-propagating front (bottom) moves much more slowly, at approximately 1.5 m/s. Within the first 5 microseconds of the dynamics, a total of 38 front positions are recorded for a virtual frame rate of nearly 8 MHz.
\label{3}}
\end{figure} %%%%%%%%%%%%%%%%%%%%%%%%%%%%%%%%%%%%%%%

\subsection*{Droplet Impact and the VFT}

The VFT has been used to enhance the temporal and spatial resolution of droplet impact dynamics. In conjunction with TIR microscopy, the position of the wetting front can be measured at high virtual frame rates\cite{Kolinski:2012fx}. Here, a water droplet falls under gravity and impacts a solid surface illuminated from beneath in total internal reflection as described elsewhere\cite{Kolinski:2014cx,Kolinski:2012fx, Kolinski:2014jf} and shown schematically in Fig \ref{3}a. Before the droplet can contact the surface it must drain the air beneath it. As the droplet approaches the surface, the air fails to drain\cite{Mandre:2009bi, Mani:2010ek, Mandre:2011hp} and instead compresses, diverting the liquid over a nanometer scale air film\cite{Kolinski:2012fx}. Measurements using the VFT show that the liquid front velocity exceeds the liquid capillary velocity. However, the liquid must flow at the smallest scales in order to make contact; this confirms that the air mediates contact at the highest impact velocities\cite{Kolinski:2012fx}. 

A typical CFS for a droplet impacting at a velocity of 3.5 m/s is shown in Fig \ref{3}b. The CFS is used to make virtual frames, as shown in Fig \ref{3}c, from which the spatio-temporal dynamics of the advancing wetting front are measured, as shown in Fig \ref{3}d. For this impact, the VFT enables us to resolve the front position at micron-scales, with nearly 40 measurements in the first 5 microseconds of the impact process. This corresponds to a virtual frame rate approaching 8 MHz \footnote{Again, we have limited the virtual frame rate such that each frame corresponds to approximately one pixel of movement of the front.}. Such data are necessary to advance our understanding of dynamic wetting processes, as highlighted by recent experimental work \cite{Kolinski:2014cx,Kolinski:2012fx, Kolinski:2014jf} and numerical calculations \cite{Sprittles:2017kv}.

\begin{figure}[ht] %%%%%%%%%%%%%%%%%%%%%%%%%%%%%%%%%%%%%%
\includegraphics[width = 1\textwidth]{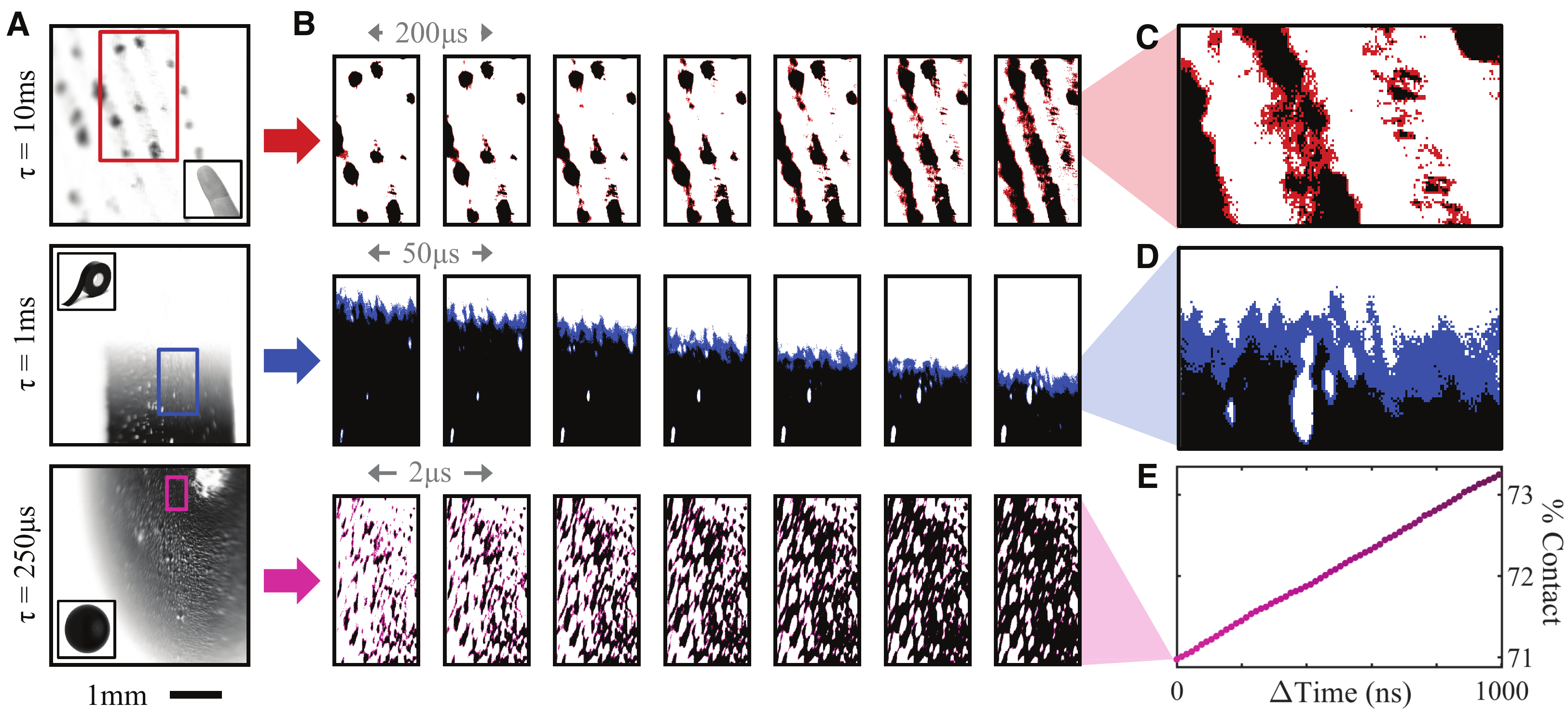}
\caption{\textbf{Time-Gated VFT with a `Slow' Camera}
a) Three experiments - an impacting finger, peeling tape, and an impacting elastic hemisphere - generate compressed frame stacks (CFS's) using a rectangular light pulse of length 10ms, 1ms, and 250$\mu$s respectively. The images are taken using the TIR lighting setup shown in Fig \ref{3}, and have a resolution of 2000x2000 pixels. b) Cropped subsections of the field of view are thresholded to generate virtual frames. Color denotes change from the previous virtual frame. c) Zoom in on fingerprint. Note the contact spreading from many small points to form the larger bands. d) Zoom in on peeling detachment front, Note the cavitation occuring ahead of the front, and the resulting rough contact line. e) Percent contact of the shown subsection over time at 65MHz. Note that this graph represents $1.5 \%$ of the field of view and $0.4\%$ of the exposure time.
\label{3b}}
\end{figure} %%%%%%%%%%%%%%%%%%%%%%%%%%%%%%%%%%%%%%%

\subsection*{Even Faster: Time-Gated VFT}

The methods described to this point may be referred to as `Continuous VFT,' because the high frame rate is mantained during continuous recording, and the virtual frames span multiple exposures. However, a further boost in frame rate is possible if continuous recording is inessential. While the boost factor $\beta$ is defined by the camera's sensor, $fps_{MAX}$ may be increased by reducing $\tau$ in equation \ref{fpsMAX}. Augmenting the frame rate in this manner can be achieved by many means, from simply reducing the electronic shutter time for cameras with a global shutter, to external gating using a pulsed light source\footnote{The shape of the light pulse can be used to arbitrarily weight the timing of the virtual frames. For example, a triangular pulse envelope results in more virtual frames at $t= \tau/2$, and fewer at $t=0$ and $t=\tau$.} or an electronically controlled filter such as an image intensifier. Using a camera capable of only 100fps with standard use, several non-trivial rapid dynamics are visualized using time-gated VFT, as shown in Fig \ref{3b}. These measurements utilize the TIR lighting configuration shown in Fig \ref{3} combined with a rectangular pulse generator powering the light source. Using a pulse length $\tau = 10ms$, a fingerprint contacting the solid surface is seen to make contact first in discrete points, which then merge to form familiar lines, as shown in Fig \ref{3b}a. It is possible that this transition influences the staying power of fingerprints left behind on a surface \cite{Merkel:2011ub}. A shorter pulse, $\tau = 1ms$, is used to observe tape being quickly peeled off of a solid surface, revealing cavitation bubbles forming \textit{ahead} of the releasing front, as noted in \cite{Villey:2015ck}, and shown in Fig \ref{3b}b. This process is not only rapid but small in comparison to the scale of the tape, requiring both high resolution and frame rate to visualize. Finally, the dynamics of an impacting hemispherical elastic solid are visualized using a pulse of length $\tau = 250\mu s$, as shown in Fig \ref{3b}c. The frame rate for this measurement exceeds 65MHz while mantaining a 4Mpx field of view.  A central bubble of air is mantained for the entire exposure, reminiscent of the dynamics of the impacting droplet in Fig \ref{2}, wherein two fronts fill the contact surface. Furthermore, in dynamics not observed in the liquid analog, initially patchy contact is established, and is filled in over $10 \mu s$ timescales. While there are successful theories that allow us to discuss contact at small scales \cite{BowdenTabor, Persson:2001kz}, as well as dynamic contact \cite{Johnson:1987vx}, they do not include the effect of air, which appears to be deforming our elastic impactor. We note that one may increase the frame rate arbitrarily by further shortening $\tau$, the only drawback being a proportionally smaller measurement time window.

\begin{figure}[ht] %%%%%%%%%%%%%%%%%%%%%%%%%%%%%%%%%%%%%%
\includegraphics[width = 1\textwidth]{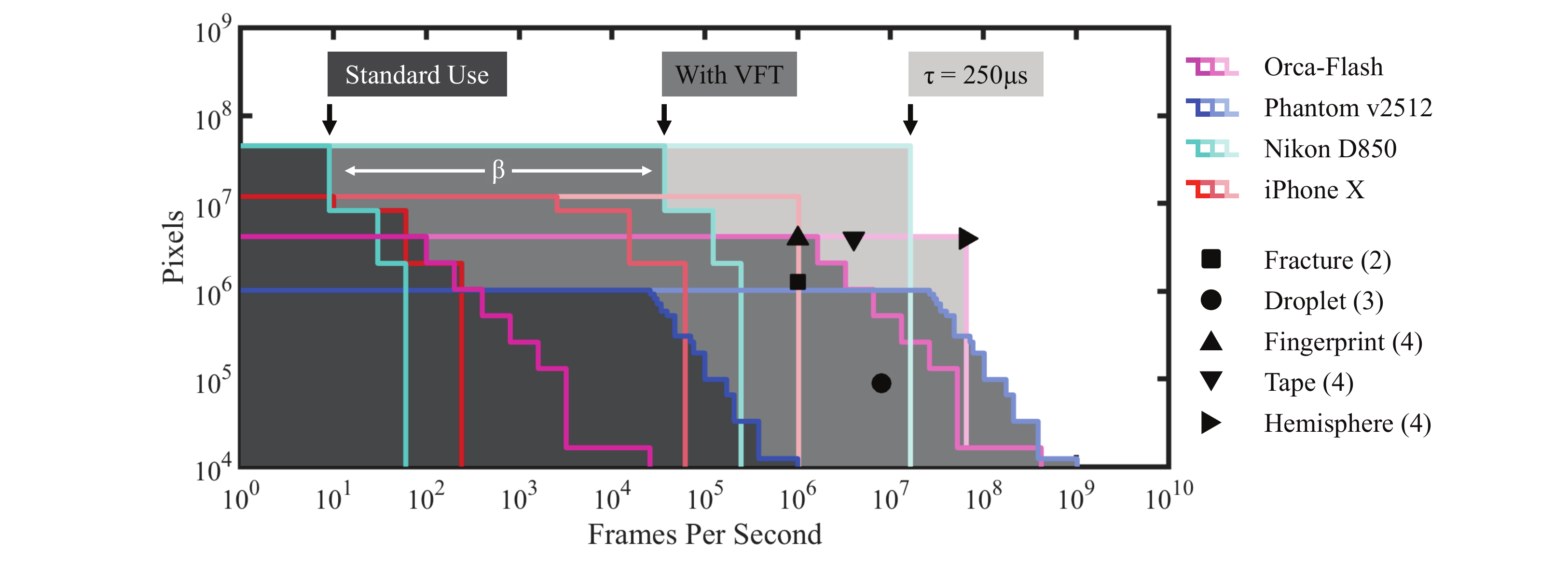}
\caption{\textbf{VFT Capability Phase Space}
Conventional fast cameras demand a tradeoff between pixel resolution and frame rate (bright lines, dark gray area on left); VFT introduces an alternative tradeoff between bit depth and frame rate resulting in an enhanced frame rate without loss of spatial resolution (faded lines, medium gray area in middle). Note that $\beta \equiv 2^{\#bits-2}$ to account for typical sensor and signal noise. This enhanced virtual frame rate does not require any change in the operation of the camera; indeed, the dynamics can be continuously recorded over many exposure times. Using Time-Gated VFT with a controlled exposure time $\tau$, the frame rate may be increased arbitrarily to $\beta/\tau$ using equation \ref{fpsMAX}, shown using the most faded lines and the lightest gray area (on right) for $\tau=250 \mu s$. Black symbols indicate experimental data from Figs \ref{2}, \ref{3}, and \ref{3b}.
\label{4}}
\end{figure} %%%%%%%%%%%%%%%%%%%%%%%%%%%%%%%%%%%%%%%

\section*{Discussion}

The capabilities of the VFT are summarized in the figure of merit, Fig \ref{4}, using the specifications of several conventional and fast cameras. Here, the frame rate is plotted as a function of the number of pixels recorded. For a state-of-the-art high speed camera such as the Phantom v2512 (blue), the trade-off between region of interest and frame rate is clear; as the read-out rate is defined in pixels-per-second, the product of frame rate and pixel number is nearly constant. Using the VFT, this high speed camera receives a maximum frame rate boost of $\beta = 2^{12-2} = 1024$, allowing sub-nanosecond frame rates even allowing for 2 bits of sensor and signal noise. The advantage of VFT is even more evident in high-bit depth cameras like the Hamamatsu Orca-Flash4.0 V3 (pink) and the Nikon D850 (teal), with bit depths 16 and 14 respectively. Note the large increase in recording capability (medium gray) resulting from using continuous VFT across all cameras. Time-gated VFT (lightest gray) can generate even faster recording speeds, allowing `slow' cameras to generate blistering virtual frame rates, with multi-megapixel resolution. As an example, the Nikon D850 using a $\tau = 250 \mu s$ light pulse records virtual frames at 16MHz, as shown by the lightest teal curve in Fig \ref{4}, while maintaining a resolution over 50Mpx! The performance of the VFT when used with the iPhone X (red) highlight the versatility and platform independence of the technique, and open the door for high-speed imaging experiments without specialized and expensive equipment.

We have described a novel and straightforward method for capturing high speed phenomena in detail in the Virtual Frame Technique. When suitable pre-processing is possible, high speed phenomena can be recorded at rates circumscribed only by the camera's bit depth and the exposure time, altering the typical modality of high-speed imaging, wherein the frame rate varies inversely with the number of pixels recording. High speed imaging typically employs short exposure times to reduce blur, capturing discrete, sharp snapshots. However using the VFT one may eliminate the gaps between these frames, recording the dynamics continuously\footnote{Continuity of recording is limited only by the reset time of the camera's sensor}. Counter-intuitively, one may record dynamics with crisper resolution in both time and space by increasing exposure time to record virtual frames.

\section*{Supplementary Video}
 \textbf{Elastomer\_Fracture\_1Mfps\_at\_30fps.mp4} (Fig \ref{2}) The VFT is tested by recording images simultaneously using a slow camera and a traditional high-speed camera. The tradeoff between the high rate and region of interest is obvious for the traditional high speed camera, whereas the greater data density at larger observed area offered by the VFT is equally apparent. Frames are replayed at 30 fps (slowed down by 33,000).

\bibliography{bib}

\section*{Acknowledgements}

J.M.K. acknowledges funding from EMSI and EPFL. 

\section*{Author contributions statement}

All authors conceived the experiment and wrote and reviewed the manuscript. S.D. conducted most of the experiments.

\end{document}